# Chapter

# Notions of Chaotic Cryptography: Sketch of a Chaos based Cryptosystem

Pellicer-Lostao Carmen, López-Ruiz Ricardo
*Department of Computer Science and BIFI,
University of Zaragoza, Spain*

## 1. Introduction

Chaotic cryptography describes the use of ***chaos theory*** (in particular physical dynamical systems working in chaotic regime as part of communication techniques and computation algorithms) to perform different cryptographic tasks in a cryptographic system.

Then, we can start by answering the questions: what exactly is chaos?, how is it used in cryptography?. First of all, let us say that there is not an universally mathematical accepted definition of the term "chaos". In general sense, it refers to some dynamical phenomena considered to be complex (lack of time and spatial order) and unpredictable (erratic). Although it was precluded by Poincaré at the end of the XIX century (Poincaré, 1890), chaos theory begins to take form in the second half of the XX century (Lorenz, 1963; Mandelbrot, 1977) after observations of the evolution of different physical systems. These systems revealed that despite of the knowledge of their evolution rules and initial conditions, their future seemed to be arbitrary and unpredictable. That opened quite a revolution in modern physics, terminating with Laplace's ideas of casual determinism (Laplace, 1825).

Chaos has been observed in nature, in weather and climate (Sneyers, 1997), population growth in ecology (May & McLean, 2007), economy (Kyrtsou & Vorlow, 2005), to mention only a few examples. It also has been observed in the laboratory in a number of systems such as electrical circuits (van der Pol & van der Mark, 1927), lasers (Casperson, 1988), chemical reactions, fluid dynamics, mechanical systems, and magneto-mechanical devices.

In essence, chaos theory studies systems that evolve in time presenting three particular properties of movement: sensitivity to initial conditions (dynamical instability), stretching and folding of the phase space (topological mixing) and aperiodic trajectories arbitrary close to an infinite set of periodic orbits (dense orbits). Chaos theory provides the means to explain chaos phenomena, and control and make use of chaotic dynamical systems.

A remarkable characteristic of chaotic systems is their capability of producing quite complex patterns of behavior from simple real systems or in simulations from low dimensional systems given by a small set of evolution equations. This quality has made them particularly useful for application in a wide variety of disciplines, such as biology, economics, engineering and others (Cambel, 1993; Kocarev et al., 2009). In these applications, chaotic systems are used to produce, simulate, assist or control different processes improving their performance or providing a more suitable output.

The use of chaos in cryptography seems quite natural, as its inherent properties connect it directly with cryptographic characteristics of *confusion* and *diffusion*. This idea is present in Shannon's works (Shannon, 1949), even earlier than the term "chaos" appeared in the scientific literature. Additionally, chaotic dynamical systems have the advantage of providing qualitatively simple mechanisms to generate deterministic pseudo randomness. For cryptography, this could be the promise of producing simpler or better randomness in terms of performance (Tenny et al., 2006).

By now, the history of chaos-based cryptography is more than two decades long. First, some works appear in the 80's (Wolfram, 1985; Guan, 1987), but it is in the 90's, when chaotic cryptography really takes off. Two papers mark this beginning (Matews, 1989) and (Pecora, 1989). The first one proposes a digital stream cipher where a signal generated from a chaotic system is used to mask the clear message. The second proposes chaos synchronization to mask the clear message with a chaotic signal at the physical level of the communication channel and to use synchronization techniques at the receiver to filter the chaotic signal. These two papers also open two different views of the application of chaos to cryptography that will be later referred as digital or analog techniques (Alvarez & Li, 2006).

Since then, the number of chaotic cryptosystems that have been proposed is too large to be covered in this chapter. The interested reader can find a complete and updated view of this field in (Kocarev, 2011). In addition, for more recent developments, some work is also needed to assess their security. Nevertheless these works have been published in journals of physics or engineering; they have revealed the potential of this field, but also a series of errors and weak points that need to be overcome. As a consequence, chaotic cryptography has been an active research field but with marginal impact in classical cryptography (Dachselt & Schwarz, 2001; Amigó, 2009).

In the end, the question is, can chaotic systems provide alternative techniques able to enhance cryptographic algorithms?. This chapter can be a worthy material so that the reader can find some guides in order to answer himself this question. Thus, the objective of this chapter is to give a general vision of what chaotic cryptography is and a comprehensive example that illustrates the main techniques used in this field. In it, the authors are intended to present a series of selected topics of special interest in this field.

In successive sections of this chapter the reader will be introduced to the following topics: fundamentals of chaos, relation between chaos and cryptography, different kinds of chaotic cryptosystems and their main characteristics, Pseudo-Random Number Generation (PRNG) based on digitized chaos, cipher design based on two dimensional chaotic maps, and the corresponding best practices and guidelines required for designing good cryptographic algorithms.

## 2. Chaos and Cryptography

This section explains the basic concepts of chaos theory. Then, the focus is set on those properties that seem relevant for its application to cryptography. In particular, first it is given a conceptual introduction to chaotic systems, to follow with a practical approach through non-linear dynamical systems. The section concludes considering how chaos theory can be useful for cryptographic applications.

### 2.1 Basic properties of chaotic systems

Chaos has been observed in numerous natural and laboratory systems (Sneyers, 1997; May, 1976; Casperson, 1988; Kyrtsou & Vorlow, 2005; van der Pol & van der Mark, 1927) covering a substantial number of scientific and engineering areas (physics, biology, meteorology, ecology, electronics, computer science and economy, among others). These phenomena, as mentioned before, show specific properties that make them complex and unpredictable.

Chaos theory deals with systems that evolve in time to a particular kind of dynamical behaviour. As this is a vast mathematical theory, the interested reader is addressed to (Robinson, 1995) for a broader introduction. In general, these systems obey a certain set of laws of evolution, and so, they are deterministic. It has to be said, that chaos occurs only in some deterministic non-linear systems. Explicitly, chaos appears when there is a sustained and disorderly-looking long-term evolution that satisfies certain mathematical criteria.

There is a set of properties that summarize the characteristics observed in chaotic systems. These are considered the mathematical criteria that define chaos. The most relevant are:

- **Dynamic instability:** also referred as butterfly effect, it is the property of sensitivity to initial conditions, where two arbitrarily closed initial conditions evolve with significantly different and divergent trajectories.(Boguta, 2011)

- **Topological mixing**: intuitively depicted as mixing coloured dyes, it means that the system will evolve in time so that any given region of states is always transformed or overlaps with any other given region. (Mahieu, 2011)

- **Aperiodicity**: the system evolves in an orbit that never repeats on itself, that is, these orbits are never periodic (Zech et al., 2011).

- **Dense periodic orbits**: it means that the system follows a dynamics that can arbitrarily closely approach every possible asymptotic state.

- **Ergodicity:** statistical measurements of the variables give similar results no matter if they are performed over time or space. Put it in another way, the dynamics shows similar statistics when measured over time or space.

- **Self-similarity:** the evolution of the system, in time or space, shows the same appearance at different scales of observation. This characteristic makes the system to appear auto-repetitive at different scales of observation (Fabre, 2011)

The references included above show interactive demonstrations of these properties. The following section introduces some notions of non-linear dynamical systems.

## 2.2 Non-Linear Dynamical Systems (NLDS)

A dynamical system is a physical phenomenon that evolves in time. In mathematical terms, the states of the system are described by a set of variables and its evolution is given by an equation and the value of the initial state. This is summarized in Eq. (1),

$$\frac{dX_i(t)}{dt} = F_i(X_j(t), \Lambda) \tag{1}$$

where $X_i(t) \in R^N$ is the coordinate $i$ of the state of the system at instant $t$, that is $X$ is an N-dimensional vector with $i,j=0,1,...N$ with $N \geq 1$, $F$ is a parametric function that describes the evolution of the system and $\Lambda$ is the vector of parameters that control the evolution of the system.

As chaotic systems only occur in non-linear dynamical systems, $F$ will be considered to be non-linear. For digital cryptographic applications we centre our attention on discrete-time NLDS. Then, a *discrete-time* NLDS is given by the following equation:

$$X_{i+1} = F(X_i, \Lambda) \tag{2}$$

The significance of the mathematical symbols in Eq. (2) is the same as in Eq. (1) but now the time $t$ is discrete. It is observed that this kind of systems is deterministic, thus the time evolution of $X$ can be calculated with $F$ and $\Lambda$ from a given initial state $X_0$. They are also *recursive*, as the next state is calculated from the previous state.

There are a series of concepts or terms that are of special interest in the study of NLDS. The first one is the *phase space* that is the subspace of $R^N$, where all possible states of the system are confined:

$$U \subset R^N \text{ and } F: U \to U, \tag{3}$$

where $N$ is the dimension of the phase space or *degree of freedom* of the system. The evolution in space of an initial state when time passes is called *orbit*. As we are considering $F$ a discrete-time function, the orbits of these systems will be a collection of real pairs of numbers:

$$(t_0, X_0), (t_1, F(X_0)), ... , (t_i, F_i(X_0)), ... \tag{4}$$

Then there is the central concept in chaos theory of *attractor*. The term attractor refers to the long-term behavior of the orbits, and it represents the region of phase space where the orbits of the system converge after the transitory. The attractor $A$ is a compact region where all orbits converge and where the system gets trapped,

$$A \subset U \quad \text{and } A = F(A). \tag{5}$$

Geometrically, an attractor can be a point, a curve, a manifold, or even a complicated set with a fractal structure known as a *strange attractor*. A brief description of them is given:

- **Fixed point,** it corresponds to a stationary state of the system.

- **Limit cycle,** which is associated with a periodic behaviour of the system. Once the system enters this attractor the states of the system repeat periodically.

- **Manifold,** where there are more than one frequency in the periodic trajectories of the system. For example, in the case of two frequencies, the attractor is a 2D-torus.

- **Strange attractor,** it is informally said to have a complex geometric shape with non-integer dimension. Any state in the attractor evolves within it and never converges to a fixed point, limit cycle or manifold. The dynamics on this attractor is normally chaotic, but there exist also strange attractors that are not chaotic. (Mahieu, 2011)

After reviewing the main characteristic of NLDS, one could define the term *chaotic system* as a NLDS that have at least a chaotic strange attractor.

NLDS are usually studied only in a qualitative and computational way, as opposed to the study of linear systems, where there is a set of analytical tools (Devaney, 1989). Different models of N-dimensional discrete-time mappings have been studied, and under certain circumstances complex behaviour in time evolution has been shown. The 1-dimensional cases have been deeper analyzed (Collet & Eckmann, 1980), cases of N=2 have also several well explored examples (Mira et al., 1996) and (López-Ruiz & Fournier-Prunaret, 2003), but as N increases, the complexity grows and less literature is found with a well documented analysis of the chaotic properties of the mapping (Fournier-Prunaret et al., 2006).

In chaotic cryptography the behaviour of the dynamical system is fully studied to assess the security of the cryptosystem. Due to the nature of NLDS, chaotic cryptography does not have analytic tools, as those that exist in classical cryptography. Then, qualitative techniques have to be applied in chaotic cryptoanalysis. In particular, the study of the sensitivity to the initial conditions and to the control parameters is considered of importance. Normally these elements are also a key part of the cryptosystem.

a) Assessing the sensitivity to initial conditions
Lyapunov exponents are used as a quantifier of the divergence of the orbits in NLDS. N exponents can be calculated for a N-dimensional system. These exponents allow us to decide if the dynamical system has sensitivity to initial conditions. A system is considered to be chaotic when it has a dense aperiodic orbit, and at least the higher of the Lyapunov exponents is positive.

The study of the value of *Lyapunov exponents* as a function of the control parameters (vector $\Lambda$ in Eq. (2)) is a matter of interest as it allows to discover windows of non chaotic behaviour of the system, where periodic patterns may appear.

b) Assessing the sensitivity to control parameters
Chaotic systems are also sensitive to the variation of the control parameters (vector $\Lambda$ in Eq. (2)). This dependence may produce that the dynamics of the system can be completely different (chaotic, periodic, divergent, etc...) depending on the values of $\Lambda$. A substantial change in the dynamics of the system due to a variation of the values of the control parameters is called a *bifurcation* (Mahieu, 2011).

The *bifurcations diagrams* are used to study the dynamics of the system as a function of the values of $\Lambda$. These unfolding diagrams allow knowing the regions of the phase space where the system displays chaotic or regular behaviour, depending on the values of the control parameters.

## 2.3 Connection between Chaos and Cryptography

As it has been said previously, the main characteristics of chaotic systems make them intuitively interesting for their application in cryptography. Here, this connection is explored in more detail.

Chaotic systems are implemented with deterministic NLDS, being able to produce the deterministic pseudo-randomness required in cryptography. In addition to that, NLDS are able to produce complex patterns of evolution. This gives to chaotic systems the algorithmic complexity required in cryptographic systems.

Now, let us examine how the inherent properties of chaos connect it directly with cryptographic characteristics of *confusion* and *diffusion* (Shannon, 1949). Referring to the properties discussed of chaotic systems, it is clear that the properties of ergodicity, auto-similarity, topological mixing are directly connected with confusion. The dynamics in the chaotic attractor is given by aperiodic orbits that generate similar statistical patterns. These patterns can be used to mask clear messages by means of substitution-like techniques.

On the other hand, diffusion is closely connected with the sensitivity that chaotic systems present to initial conditions and control parameters. Diffusion produces the avalanche effect where a minimum difference in the input of the cryptosystem gives a completely different output. A chaotic system produces this behaviour when a small change is applied to its initial conditions or control parameters. The use of these variables as input in the cryptosystem algorithm may produce the same avalanche effect.

Table 1 obtained from (Alvarez & Li, 2006), summarizes the connection between chaos and cryptography.

| Chaotic characteristic | Cryptographic property | Description |
| --- | --- | --- |
| Ergodicity<br>Mixing property<br>Auto-similarity | Confusion | The output of the system seems similar for any intput. |
| Sensitivity to initial conditions and control parameters | Difusion | A small diference in the input produces a very different output |
| Deterministic | Deterministic pseudorandomness | A deterministic procedure that produces pseudo-randomness |
| Complexity | Algorithmic complexity | A simple algorithm that produces highly complex outputs |

Table 1. Comparisons of chaotic and cryptographic properties (from Alvarez & Li, 2006).

To conclude, it would be interesting to describe particular potential advantages that chaotic cryptosystems may provide to cryptography (Tenny et al., 2006). First of all, chaotic systems appear spontaneously in nature and can be directly applied to security processes, as it is the case of physical devices used in communications. Those show naturally non-linear and chaotic behaviours that can be straightforwardly used to secure communications.

Also chaotic NLDS have the advantage of being implemented with simple computable deterministic algorithms. Additionally these algorithms may refer to N-dimensional equations or to several systems combined at a time. They could be subject of implantation with parallel computing and become faster algorithms.

## 3. Chaotic based Cryptosystems

This section presents two specific examples of chaotic based cryptosystems. Full versions of them can be found in (Pellicer-Lostao & López-Ruiz, 2008, 2009). These are specially suited to illustrate in detail some relevant techniques used in chaotic cryptography. In particular, they cover three interesting topics: Pseudo-Random Number Generation (PRNG) based on digitized chaos, cipher design based on two-dimensional chaotic maps and the role of symmetry and geometry of chaotic maps in encryption algorithms.

Along with that, an introduction provides a global (though not extensive) picture of different techniques used in chaotic cryptography.

### 3.1 Review of chaos based encryption techniques

The number of chaotic cryptosystems proposed in literature is too large and diverse to be covered in this chapter. Valuable overviews with corresponding references can be found in (Li, 2003; Kocarev, 2011). It is important as well, to have in mind that it is an active area of research and recent developments need some time to assess their security.

To keep this review sufficiently small and clear, it will only cover a general view of the main techniques and their basic concepts of operation. See Table 2 for a brief summary of them obtained from (Li, 2003). To conclude and provide a comprehensive background, a detailed description is dedicated to the specific technique used in the examples described later in this section.

| CATEGORY | METHOD | | DESCRIPTION |
|---|---|---|---|
| Analog cryptosystems | Additive chaos masking | | A chaotic signal is added to the message. |
| | Chaotic shift keying | | A digital message signal switches among different chaotic systems to be added to the message |
| | Chaotic modulation | | A message signal is used to change the parameters or the phase space of the chaotic transmitter |
| | Chaotic Control | | A message signal is ciphered in a classical way and used to perturbate the chaotic system |
| Digital cryptosystems | Stream ciphers | Chaotic PRNG | A chaotic signal generates a pseudorandom sequence (keystream) to XORed the message |
| | | Chaotic Inverse System approach | A message signal is added to the output of the chaotic signal, which has been fed by the ciphered message signal in previous instants |
| | Block ciphers | Backwards iterative | A block of a clear message is ciphered using of inverse chaotic systems |
| | | Forwards iterative | A block of ciphered message is obtained by pseudoramdom permutations obtained from a chaotic system |
| | | S-Boxes | An S-Box is created from the chaotic system. There can be dynamic or static S-Boxes |
| | Miscellaneous | Searching based chaotic ciphers | A table of characters is generated from a chaotic system. The table is used to cipher the characters of the message text |
| | | Cell. Automata | The chaotic system is a Cellular Automata |

Table 2. Different kinds of chaos based cryptosystems presented in literature (Li, 2003).

As it is seen in Table 2, the application of chaotic systems to cryptography has followed two main approaches. These have been called, analogue and digital techniques (Alvarez & Li, 2006). Let us discuss in further detail their principal concepts and features of operation.

a) Analogue chaos-based cryptosystems
These systems are mostly used to secure communications and have attracted very extensive research activity lately (Larson et al., 2006). In them, the clear message is masked with a chaotic signal at physical level of the communication channel. The natural non-linearity of electric and optical communication devices is controlled to produce a chaotic waveform that modulates the message in a secure way for transmission. At the receiver, chaotic synchronization techniques demodulate the signal and produce the clear message. Though no completely secure their implementation results straightforward.

Main techniques of chaotic modulation (Yang, 2004), described in Table 2, are shown in Fig.1

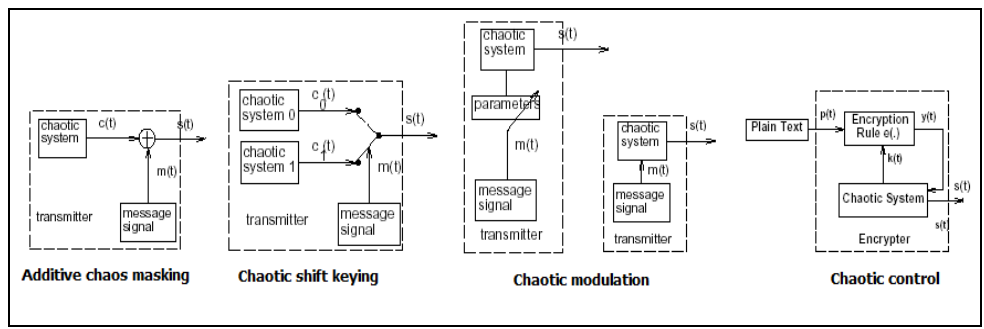

Fig. 1. Modulation techniques used in analogue chaotic cryptosystems (Yang, 2004).

Typically used in optical communications with high rates of transmission, synchronization techniques become critical, also due to the fact that chaotic performance implies highly sensitivity to small deviations. Here, a chaotic control signal is used to provide a synchronous performance of transmitter and receiver. Though not necessary to be equal (Femat et al., 2005), they require careful fabrication and operation within specific conditions.

These systems are not cryptographically secure (Larson et al., 2006), due intrinsic limitations of their design. As it has been seen in previous section, chaotic signals are deterministic and though they appear to be random, they are intrinsically correlated. This means that patterns can be found in the communication signals that may allow an attacker to decipher the message. Even though, their technical complexity and high transmission rates makes them valuable for communications that only require security for a limited period of time.

b) Digital chaos-based cryptosystems
These cryptosystems are basically algorithms implemented in digital circuits or computers. These algorithms are based in iterative computations of chaotic functions that produce digital signals. Then, basic cryptographic operations (substitution and mixing) are used to mask the clear message with the chaotic signals. These cryptosystems involve one or more chaotic systems in the algorithm and use their initial conditions and/or control parameters as secret keys.

But chaos implemented on computers with finite precision may diverge from real-precision chaos. This is why it is normally called "pseudo chaos". In pseudo chaos, the chaotic properties of the system may suffer dynamical degradation, for pseudo orbits may depart from the real ones in many manners (Guckenheimer & Holmes, 1983; Li, 2003; Li et al, 2005). Even so, it is possible for these pseudo chaotic systems, to minimize their dynamical degradation. In this direction, the idea of using high dimensional chaotic systems may also be helpful. While less known, these systems whirl many variables at any calculation and the periodic patterns produced by the finite precision are diminished (Falcioni et al., 2005). Anyway, only a detailed study of the dynamical system may guarantee their performance.

Main techniques used here are algorithms that implement stream or block ciphers (Li, 2003). Understanding by them, the encryption of individual bits or blocks at a time, respectively. There are also miscellaneous systems that use characteristic techniques or systems and belong to specific families. As for security, cryptographers demand more rigorous studies of security in publications (Alvarez & Li, 2006). Also the novelty of chaotic cryptography shows a lack of tools for cryptanalysis as the ones of classical cryptography (Amigó, 2009).

c) Digital stream ciphers implemented with chaotic PRBG

Pseudo Random Bit Generation (PRBG) is a topic of high interest. As it is shown in Table 2, these are used in stream cipher encryption but they are also a subject of broad application in many scientific and engineering areas. This is why, since 1990 the proposals of chaos based PRBG have demonstrated great development (Kovarev, 2001; Kocarev & Jakimoski, 2003).

To build a chaotic PRBG, we take an N-dimensional deterministic discrete-time dynamical system. As seen in Eq. (2), it is an iterative map $f: R^N \to R^N$ of the form:

$$X_{k+1} = F(X_k, \Lambda) \qquad (6)$$

where $k = 0, 1, ..., n$. is the discrete time, $\Lambda$ vector of parameters, $X_0$ initial condition and $X_1, ..., X_n$ states of the system in the following instants of time. Then it is necessary to construct a numerical algorithm that transforms the states of the system into binary numbers. Additionally not all states of the orbit are used to produce the pseudo-random sequence. Normally the orbit is sampled in order to get rid of the correlation existent between consecutive states. The existing designs of chaotic PRBGs use different techniques to pass from the continuum to the binary world (Li, 2003). The most relevant are:

- **Extracting bits** from each state along the chaotic orbits (Protopopescu, 1995; Bodgan et al., 2007; Fournier-Prunaret & Abdel-Kaddous, 2012).

- **Dividing the phase space into m sub-spaces**, and output a binary number $i = 0, 1,...,m$ if the chaotic orbit visits the i-th subspace (Stojanovski & Kocarev, 2001; Suneel, 2009).

- **Combining the outputs of two or more chaotic systems** to generate the pseudo-random numbers (Li et al, 2001; Po et al., 2003).

The binary sequence generated with the chaotic PRBG algorithm is used as a keystream. To mask the clear message, the keystream is added to it through a binary XOR operation. The initial conditions and the vector of parameters are used as the secret key.

## 3.2 First example: Building a stream cipher based on 2D chaotic maps

This subsection presents an asymmetric or secret key chaotic cryptosystem. It describes a chaotic digital stream cipher. Its nucleus is a Pseudo-Random Bit Generation (PRBG) based on 2D chaotic mappings of logistic type (Pellicer-Lostao & López-Ruiz, 2008). This chaotic PRBG produces pseudorandom binary sequences out of the chaotic dynamics of the considered maps. The sequences produced are used as the keystream in the cipher.

To illustrate its implementation and functionality, the design of the chaotic PRBG algorithm is fully described and different evaluations are presented, such as statistical tests, predictability and speed measurements. Considering the application of this PRBG in cryptography, the size of the available key space is also calculated.

First let us see the chaotic systems that are going to be used in this example. In (López-Ruiz & Pérez-García, 1991) the authors analyze a family of three chaotic systems obtained by coupling two logistic maps. The focus here will be made on models (a) and (b), which will be referred as System A and B:

$$\begin{array}{ll} SISTEM\ A: & SISTEM\ B: \\ T_A: [0\ 1] \times [0\ 1] \to [0\ 1] \times [0\ 1] & T_B: [0\ 1] \times [0\ 1] \to [0\ 1] \times [0\ 1] \\ x_{n+1} = \lambda(3y_n+1)x_n(1-x_n) & x_{n+1} = \lambda(3x_n+1)y_n(1-y_n) \\ y_{n+1} = \lambda(3x_n+1)y_n(1-y_n) & y_{n+1} = \lambda(3y_n+1)x_n(1-x_n) \end{array} \quad (7)$$

The interest of using these systems is their symmetry properties ($T_A(x,y) = T_B(y,x)$ and $T^2_A(x,y) = T^2_B(x,y)$) wich will be of used in 3.3 to improve the algorithm. From a geometrical point of view, both present the same chaotic attractor in the interval $\lambda \in$ *[1.032 1.0843]*. The parameter $\lambda$ in Eq. (2), could be considered different for the calculus of $x$ and $y$ coordinates. This would give two independent control parameters $\lambda_x$ and $\lambda_y$, which in a similar rank *[1.032, 1.0843]* maintain the system in a chaotic regime due to its structural stability. The dynamics in this regime (Pellicer-Lostao & López-Ruiz, 2011a) is particularly interlaced around the saddle point *P4*:

$$P4 = [P4_x,\ P4_y] \qquad P4_x = P4_y = \frac{1}{3}\left(1 + \sqrt{4 - \frac{3}{\lambda}}\right) \qquad (8)$$

Despite of the similarities of these systems, their dynamics have some differences. In Fig. 2 one can see one orbit and its spectrum for both systems with equal initial conditions.

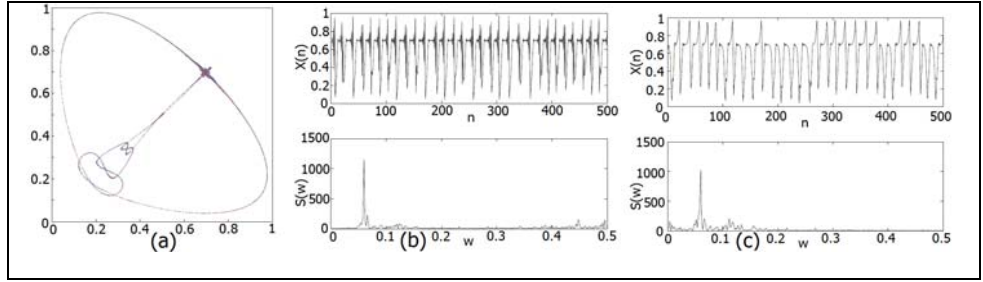

Fig. 2. Chaotic Attractor with $\lambda = 1.07$, $X_0 = [0.737, 0.747]$ for System A in (a). Orbit $x(n)$ and spectrum $S(w)$ for and Systems A and B in (b) and (c) respectively for the same $\lambda$ and $X_0$.

Second, let us present the algorithm used to obtain the Symmetric Coupled Logistic Map PRBG. In (Suneel, 2009) the author presents an algorithm to build a PRBG based on a 2D chaotic system, the Hénon map (Hénon, 1976; Pellicer-Lostao & López-Ruiz, 2011b). This algorithm is based on the technique of subspace division. The pseudorandom sequences generated with this algorithm show good random properties when subjected to different statistical tests. The author indicates that the choice of the Hénon map is rather arbitrary and similar results should also be attainable with other 2D maps. This algorithm is then taken and the chaotic Hénon map is substituted by the Symmetric Coupled Logistic Map.

But, as it normally happens in chaotic cryptosystems, this algorithm is dependent of the chaotic system and the mere substitution of the chaotic systems is not enough to maintain its performance (Li, 2003). In particular, it is found that the sub-space division method has also to be adapted to fit the geometrical characteristics of the new chaotic systems. In (Pellicer-Lostao & López-Ruiz, 2008) the interested reader can find these details. The final algorithm applied to the Symmetric Coupled Logistic Maps is shown in Fig.3 through a functional diagram. In the following paragraphs the details of its implementation are described.

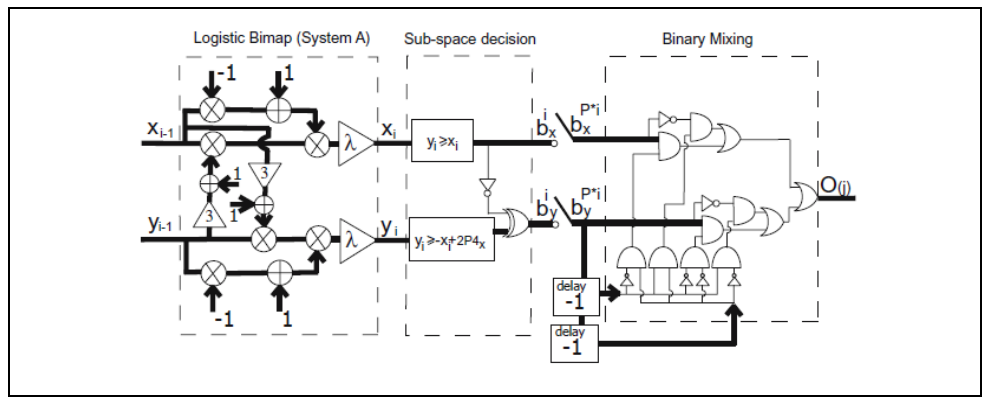

Fig. 3. Functional block structure of the proposed algorithm (example for System A)

The input of the algorithm are a state of initial conditions $X_0 = [x_0, y_0]$ and a value of the parameters ($\lambda_x$, $\lambda_y$). The functional block named as **Logistic Bimap** produces a series of iterations through Eq.(4). Whenever the initial input values produce chaotic regime this block produces a chaotic sequence of 2D real states $X_i = [x_i, y_i]$. Then, it is used the technique of dividing the phase space in four sub-spaces to transform the 2D states into binary vectors $[b^i_X, b^i_Y]$. This is done in the block named as **sub-space decision**. This produces two binary sequences $S_X = \{b^i_X\}^\infty_{i=1}$ and $S_Y = \{b^i_Y\}^\infty_{i=1}$.

As it is said before, the division of the phase space in four sub-spaces must be defined, in a way that the Logistic Bimap system visits each subspace as the Hénon map does in (Suneel, 2009). A finite automata that summarizes the pattern of visits is used to obtain that. A finite automata is inferred from the Hénon map in (Suneel, 2009) and a division of sub-paces is chosen in the Logistic Bimaps in order to get the same finite automata. In particular, if we name the sub-spaces corresponding to $[b^i_X, b^i_Y]$ with values *[0, 0]*, *[1, 0]*, *[0, 1]* and *[1, 1]* as *1*,*2*,*3* and *4*, we can analyze the detail of this pattern of visits. This is shown in Fig.4.

In (Suneel, 2009) the Hénon map does not visit the four sub-spaces equally. It is observed that there exists a symmetry of movements between sub-spaces *1-3* and *2-4*, which has a characteristic mixing of *50%* and *50%*, as long as a predominant (*80%*) and constant transition between *3* and *2*. This leads to a highly variation of binary values in sequences $S_x$, $S_y$. In the end, these conditions give the final result of an output sequence $O(j)$ with a proper balance of zeros and ones, or put it in another way, with pseudo-random properties.

To get this automata for the symmetric coupled logistic maps Systems A and B, one should chose the diagonal axis, which divides phase space in two parts, each of which is equally visited (*50%*). And additional statistical calculus is required to divide these two sub-spaces, in another two with a visiting rate of *40%-10%* each one. When this is done, one can observe that this is got by merely selecting *P4* and the line perpendicular to the axis in *P4* as the other division line.

The final sub-space division for each system is presented in Fig. 4(a) and 4(c), along with the indications of the evolution of the visits to each sub-space. For Systems A and B, this finite automata is depicted in Fig. 4(b) and 4(d).

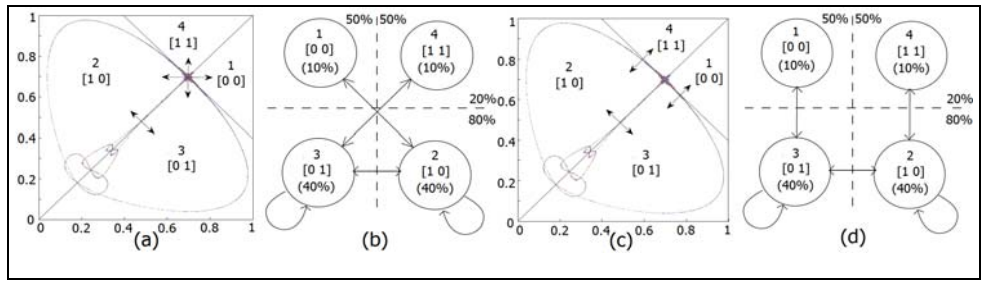

Fig. 4. (a) Final sub-space division and (b) finite automata for System A. (c) Final sub-space division and (d) finite automata for System B. (In both cases, $\lambda = 1.07$).

Finally the operation performed in the sub-space decision block is summarized in Eq (5):

$$b_x = \begin{cases} 0 & si \quad y < x \\ 1 & si \quad y \geq x \end{cases} \qquad b_y = \begin{cases} 0 & si \quad (y < x \ \& \ y \geq -x + 2P4_x) OR (y \geq x \ \& \ y < -x + 2P4_x) \\ 1 & si \quad (y < x \ \& \ y < -x + 2P4_x) OR (y \geq x \ \& \ y \geq -x + 2P4_x) \end{cases} \qquad (5)$$

After obtaining $S_X = \{b^i{}_X\}^\infty_{i=1}$ and $S_Y = \{b^i{}_Y\}^\infty_{i=1}$, they are sampled with a frequency of *1/P* (each *P* iterations) and $B_X = \{b^{P*i}{}_X\}^\infty_{i=1}$ and $B_Y = \{b^{P*i}{}_Y\}^\infty_{i=1}$ are obtained. The effect of skipping *P* consecutive values of the orbit is necessary to get a random macroscopic behaviour. With this operation, the correlation existing between consecutive values generated by the chaotic system is eliminated.

This is done in a way that over a $P_{min}$, sequences generated with $P > P_{min}$ will appear macroscopically random. Although *P* is normally introduced as an additional key parameter in pseudo-random sequences generation (Kocarev, 2001), it strongly determines the speed of the generation algorithm, so it is recommended to be kept as small as possible.

The output binary pseudorandom sequence $O(j)$ is obtained by a mixing operation of the actual and previous values of the sequence $B(j) = [B_X(j);B_Y(j)]$ given by the truth table sketched in Table 3.

|  | $B_Y(j-1)$ |  |
|---|---|---|
| $B_Y(j-2)$ | 0 | 1 |
| 0 | $B_X(j)$ | $Not(B_X(j))$ |
| 1 | $B_Y(j)$ | $Not(B_Y(j))$ |

Table 3. Truth table generating the binary sequence $O(j)$.

As a direct application for cryptography, the above described PRBG could be used in the construction of a stream cipher. Different initial conditions, $x_0$ and $y_0$ and parameters $\lambda$ and $P$, can be applied to the input of the system and be used as a key to generate the keystream, or output sequence $O(j)$. The keystream $O(j)$ can be XORed directly with a clear text, obtaining that way the ciphertext.

In rest of this example, we evaluate the functionality of the chaotic PRBG with the calculus of the size of the key space, applying statistical and predictability tests to different sequences produced with it and measuring its speed.

a) Calculus of the key space

Let us determine the operative range of initial conditions and parameters values that can be applied to the PRBG in Fig. 3. This range, when the PRBG is used in cryptography applications is known as the key-space. Then, the key space is determined by the interval of the parameter and the initial conditions that keep the dynamical system in the chaotic regime. These are $\lambda \in$ *[1.032, 1.0843]*, $x_0 \in$ *(0, 1)* and $y_0 \in$*(0, 1)*. The parameter $\lambda$ can be considered as two independent parameters $\lambda_x$ y $\lambda_y$, and the dynamic properties of the system still remain chaotic in the same interval. The sampling parameter can also be considered as another parameter of the key space. One must observe that $P$ should be kept in a suitable range, so that the PRBG is fast enough for its desired application. These intervals can be denoted with brackets and calculated as *[$\lambda_x$]= [$\lambda_y$]=0.0523, [$x_0$]=1, [$y_0$]=1* y *[P]=890,* when taking *[P]$\in$[110,1000]* as the functional range of the sampling factor.

The size of the key space will be determined by the numeric representation used in the calculus of the algorithm. If floating-point representation under standard IEEE 754 is chosen, the smallest available precision is $\varepsilon_{32} \approx$*1.1921$\times$10$^{-7}$* for simple precision with 32 bits and $\varepsilon_{64} \approx$*2.2204$\times$10$^{-16}$* for double precision with 64 bits. These quantities give the maximum number of possible values of every parameter in any of the two representations. This is easily computed dividing the intervals by $\varepsilon$, as $K_{\lambda x}=[\lambda_x]/\varepsilon$, $K_{\lambda y}=[\lambda_y]/\varepsilon$, $K_{x0}=[x_0]/\varepsilon$, $K_{y0}=[y_0]/\varepsilon$, and $K_P=[P]$. Notice that the calculus of $K_P$ is different as $P$ can only take integer values.

The total size of possible parameter values is given by $K$, calculated as $K=K_{\lambda x} \times K_{\lambda y} \times K_{x0} \times K_{y0} \times K_P$. $K$ is the size of the available key-space and its logarithm in base 2 gives us the available length of binary keys or entries to produce pseudo-random sequences in the generator. The values obtained for each number precision, are $K_{32}$=*1.21$\times$10$^{28}$*, with a key length of *93* bits for single precision and $K_{64}$=*1.00$\times$10$^{63}$*, with a key length of *209* bits for double precision.

These results are fine for the use of the chaotic PRBG in cryptography, where a length of keys greater than *100* is considered strong enough against brute force attacks, (Alvarez & Li, 2006). Nevertheless, it has to be said for accuracy's sake of that the calculus of the key space is a coarse estimation and a deeper study is required (Alvarez & Li, 2006) for accurate evaluation. The high sensitivity to parameter values and initial conditions of chaotic systems may produce windows on periodicity in the key space. Another possibility is that the dynamics can diverge towards infinity. In the systems presented here an initial calculus of *100* iterations is enough to ensure the boundless or goodness of the initial conditions.

a) Statistical testing

In general, randomness cannot be mathematically proved. Alternatively, different statistical batteries of tests have been proposed over time to assess randomness. Each of these tests evaluates a relevant random property expected in a true random generator. Then, to test a certain randomness property in a PRBG, several output sequences are taken from the generator. As one knows a priori the statistical distribution of possible values that true random sequences would be likely to exhibit for that property, a conclusion can be obtained upon the probability of the tested sequences to be random.

There exist different well-known sources of test suites available in literature, such as those described by Knuth (Knuth, 1997), the Marsaglias Diehard test suite (Marsaglia, 1995) or those of the National Institute of Standards and Technology (NIST) (Rukhin et al., 2010). But there are many more (Rütti et al, 2004; Mascagni, 1999), perhaps not so nicely packaged, but still useful. In these collections of tests, each test tries a different random property and gives a way of interpreting its results. In this example, the Diehard test suite (Marsaglia, 1995) and the NIST Test Suite (Rukhin et al., 2010) were selected, for they are very accessible and widely used. Table 4 lists the tests comprised in these suites.

| $N^{er}$ | *Diehard test suite* | *NIST test suite* |
|---|---|---|
| 1 | Birthday spacings | Frequency (monobit) |
| 2 | Overlapping 5-permutation | Frequency test within a block |
| 3 | Binary rank test | Cumulative sums |
| 4 | Bitstream | Runs |
| 5 | OPSO | Longest run of ones in a block |
| 6 | OQSO | Binary matrix rank |
| 7 | DNA | Discrete fourier transform |
| 8 | Count-the-1's test | Non-overlapping template matching |
| 9 | A parking lot | Overlapping template matching |
| 10 | Minimum distance | Maurer's universal statistical |
| 11 | 3D-spheres | Approximate entropy |
| 12 | Squeeze | Random excursions |
| 13 | Overlapping sums | Random excursions variant |
| 14 | Runs | Serial |
| 15 | Craps | Linear complexity |

Table 4. List of tests comprised in the Diehard and NIST test suites.

In each test, a *p-value* is obtained. This value summaries the strength of evidence against the randomness of the tested sequence. In Marsaglia's Diehard test suite, *p-values* should lie within the interval *[0, 1)* to accept the PRBG. In NIST Statistical Test Suite, the acceptance happens when the *p-values* are greater than *a,* the significance level of the test.

To assess the randomness of the PRBG obtained in the previous section with systems A and B, several sequences are obtained and submitted to the Diehard and NIST test suites described in Table 4. The significance level of the tests was set to a value appropriate for cryptographic applications (*a*= 0.01). Similar results were found for both systems and for simplicity, only those obtained with system A will be presented here after. Ten sequences were generated with six different sets of initial conditions and considering $\lambda = \lambda_x = \lambda_y$. Their characteristics are described in Table 5.

| *Sequence* | *S1* | *S2* | *S3* | *S4* | *S5* | *S6* |
|---|---|---|---|---|---|---|
| $x_0$ | 0.9891 | 0.4913 | 0.6727 | 0.7268 | 0.3956 | 0.9998 |
| $y_0$ | 0.6891 | 0.6913 | 0.4977 | 0.9018 | 0.4956 | 0.6498 |
| $\lambda$ | 1.048 | 1.053 | 1.069 | 1.080 | 1.064 | 1.074 |
| $P_{Dmin}$ | 55 | 45 | 35 | 47 | n.a. | n.a. |
| $P_{Nmin}$ | 83 | 105 | 83 | 83 | 100 | 85 |

Table 5. Parameters $P_{Dmin}$ and $P_{Nmin}$ for different sequences $S_i$, *i = 1,…,6,* with different initial conditions *($x_0$, $y_0$)* and control parameter $\lambda$.

Six of them (*S1, S2, S3, S4, S5* and *S6*) were tested with Nist tests suite with **200 Mill**. of bits and four of them (*S1, S2, S3* and *S4*) were tested with Diehard tests suite with **80 Mill**. of bits. Here, the parameters $P_{Dmin}$ and $P_{Nmin}$ are the minimum sampling rate or shift factor, $P_{min}$, over which, all sequences generated with the same initial conditions and $P > P_{min}$ pass Diehard or Nist tests suites, respectively. It is observed here, that the Nist tests suite requires a higher value of $P_{min}$ and that *S5* and *S6* were not tested with Diehard battery of tests.

In the Diehard tests suite, each of the tests returns one or several p-values which should be uniform in the interval *[0,1)* when the input sequence contains truly independent random bits. The software available in (Marsaglia, 1995) provides a total of **218** p-values for **15** tests, and the uniformity requirement can be assessed graphically, when plotting them in the interval *[0,1)*. For example Fig. 5 shows the *p-values* obtained for three sequences (a),(b) and (c) with the same initial conditions *S1*, and different sampling factor *P*. The first one, Fig. 5(a), demonstrates graphically the failure of the tests, for there is a non-uniform clustering of p-values around the value *1.0*. Fig. 5(b) shows the uniformity obtained with $P_{Dmin}$= **55** over the interval *[0,1)*. A better uniformity can be appreciated when $P > P_{Dmin}$ in Fig. 5(c). Sequences *S1* to *S4* where proved to pass the Diehard battery of tests with significance level *a= 0.01*. Fig. 5(d) presents a graphical representation of the *p-values* obtained for each sequence with sampling factor $P=P_{Dmin}$ of Table 6. It can be observed that some *p-values* are occasionally near *0* or *1*. Although it can not be well appreciated in the figure, it has to be said that those never really reach these values.

In the Nist tests suite, one or more p-values are also returned for each sequence under test. These values should be greater than the significance level *a*, which was selected as *a=0.01* (as in the Diehard case). These tests also require a sufficiently high length of sequences and to prove randomness in one test, two conditions should be verified. First, a minimum percentage of sequences should pass the test and second, the p-values of all sequences should also be uniformly distributed in the interval *(0, 1)*. For this case, each of the six sequences with initial conditions *S1* to *S6* are arranged in **200** sub-sequences of *1Mill*. bits each and submitted to the Nist battery of tests. Sequences $S_i$ proved to pass all tests over a minimum value $P_{Nmin}$, shown in Table 5.

In Fig.5(e) and 5(f), the results obtained for *S1* and *S4* respectively are graphically presented, as an example of what was obtained for each *Si*. The tests in the suite are numbered according to Table 4. Fig. 5(e) represents the percentage of the *200* sub-sequences of *S1* that pass each of the *15* tests of the suite. These percentages are over the minimum pass rate required of *96.8893%* for a sample size of *200* binary sub-sequences.

Fig. 5(f) describes the uniformity of the distribution of *p-values* obtained for the *15* tests of the suite. Here, uniformity is assessed. The interval *(0, 1)* is divided in ten subintervals (*C1,C2,...,C10*) and the number of *p-values* that lay in each sub-interval, among a total of *200*, are counted and proved to be uniform.

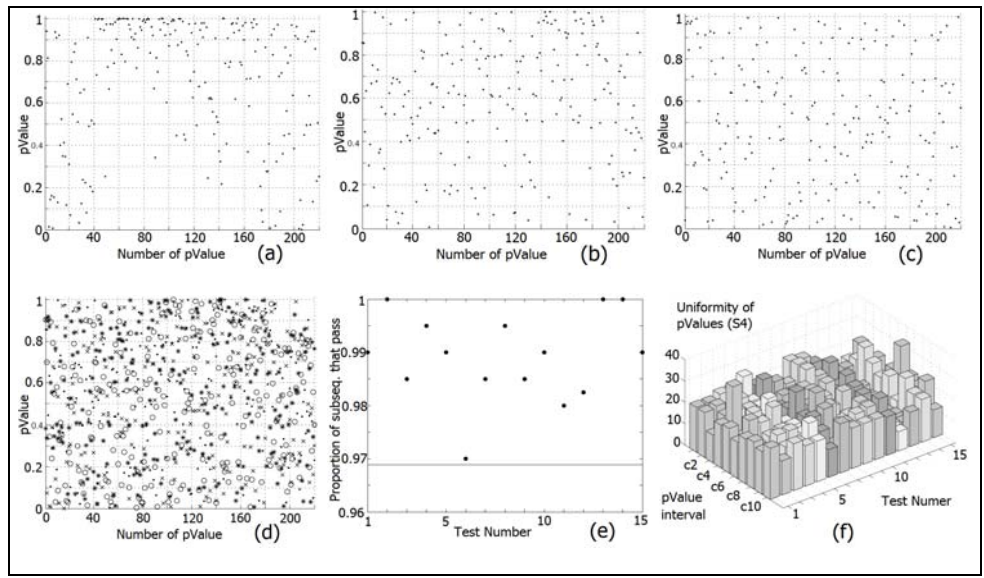

Fig. 5. Diehard test suite *p-values* obtained for initial conditions *S1* with (a) *P = 20*, (b) *P = $P_{Dmin}$= 55* and (c) *P = 110*. In (d), *p-values* obtained for initial conditions *S1, S2, S3* and *S4* with *P= $P_{Dmin}$* of Table 5. NIST test results, in (e) the proportion of sub-sequences of *S1* that passes each test, in (f) the distribution of *p-values* of *S4* is examined for each test to ensure uniformity.

The results above obtained are finally compared with the ones produced by other standard PRBG. To do that, two generators provided in the software packet of Diehard (Marsaglia, 1995) are submitted to the statistical tests.

The first one is the shift-register, generator number *13* with parameters *L1= 13, R=17* and *L2=5*. The second one is a extended congruential generator; the first one provided in the packet with seeds *78, 29* and *33*. These generators produced sequences two sequences of *11 Mill.* of bytes, more than *80 Mill.* bits. Another two generators given in NIST (Rukhin et al., 2010) were also used to produce another two standard pseudorandom sequences. In particular two sequences of *200 Mill.* bits were produced with the ANSI X9.17(3-DES) and the linear congruential generator.

The results obtained for these standard generators are similar to the ones obtained with the Symmetric Coupled Logistic Map PRBG and they are presented in Fig.6. Only one of the standard generators shows poorer performance than the chaotic PRBG. This is the shift-register generator that fails with *p-value=1* the following tests: binary rank test and count-the-1's test.

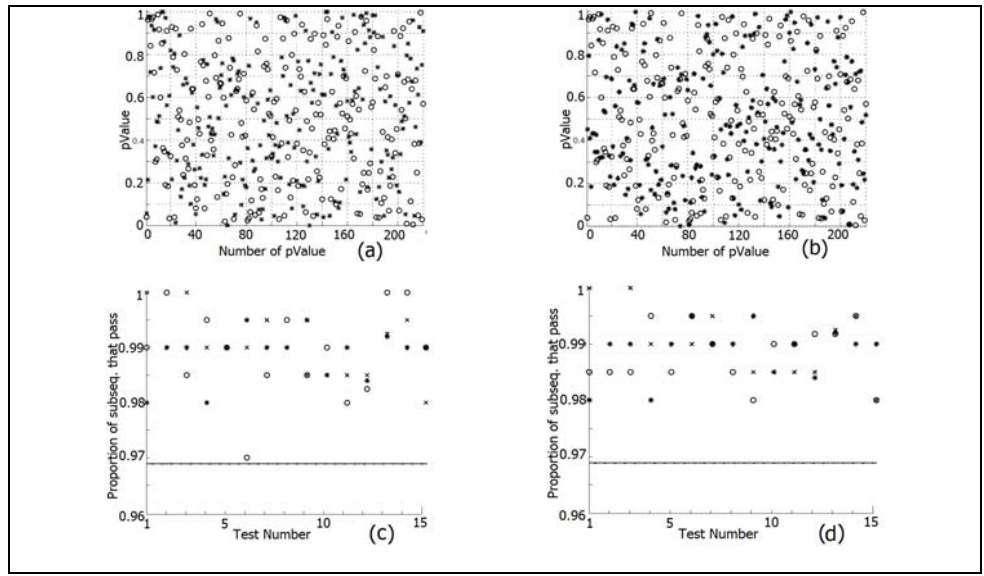

Fig. 6. Diehard test suite *p-values* obtained sequence generated with the chaotic PRBG and initial conditions *S1* and *P=110* (circles) and for the sequence generated with the standard generators (stars). (a) The shift register. (b) The extended congruential generator. (c) Proportion of sub-sequences that passes each one of NIST tests with generator ANSI X9.17 (black cross) and linear congruential generator (asterisk) and chaotic PRBG (circles) with initial conditions *S1* and *P=50* (c) or *P=90* (d).

c) Predictability testing

A PRBG is called a cryptographically secure pseudorandom bit generator (CSPRBG) when is not predictable. [Menezes et al, 1996]. To assess the unpredictability it is necessary to pass any of the following tests: the polynomial-time statistical test or the next-bit test. Both tests are equivalent.

A PRBG passes the first one if no polynomial-time algorithm can correctly distinguish between an output sequence of the generator and a truly random sequence of the same length with probability significantly greater than *0.5*. A PRBG passes the next-bit test if there is no polynomial-time algorithm which, on input of the first *L* bits of a generated sequence *s*, can predict the *(L + 1)*st bit of *s* with probability negligible greater than *0.5* (a function is negligible if it is eventually smaller than the inverse of any positive polynomial). As it is not possible to prove the non existence of something, a generator is considered to pass these tests if it is possible to verify the above conditions under some plausible but unproved mathematical assumption.

Conventional cryptography has a framework of concepts and techniques that allows conclude on predictability, based on number and computation theories. In chaotic cryptography one may expect predictability to be a function of the dynamical parameters of the chaotic system. However, here there is a lack of theoretical tools to carry out such kind of analysis. In return, one can perform empirical measurements.

This part presents several measurements to assess the predictability of the PRBG. These will show, that due to the characteristics of the design of the PRBG, the probability of producing a *0* or a *1* at a given instant *t=j* is:

$$P(O(j)=1) = P(O(j)=0) = 0.5+\Delta\varepsilon \tag{6}$$

and the maximum value of *Δε* will be measured (*Δε$_{max}$*).

To do that, the chaotic PRBG algorithm is revised. The conditions necessary for unpredictability are analyzed. After that, a binary pseudorandom sequence is generated, the probability of obtaining a *0* or a *1* bits is measured on that sequence and the error magnitude *Δε* is evaluated. Our discussion is centred in System A, but measurements could also be performed to System B the same way and similar results should be obtained.

The design of the generator is depicted in Fig.3. There it can be seen that the output bit generated at instant *t=j*, *O(j)* is generated by means of Table 4 as a function of the binary vectors *B$_Y$(j-1)* , *B$_Y$(j-2)* y *B(j) = [B$_X$(j),B$_Y$(j)]* where at instant *t=j=P*i*, *B$_X$(j)= b$^{P*i}_X$* y *B$_Y$(j)= b$^{P*i}_Y$* . As a result the output bit is produced explicitly as it is shown in following Table 7.

|  | [B$_X$(j),B$_Y$(j)] | | | |
| --- | --- | --- | --- | --- |
| [B$_Y$(j-2),B$_Y$(j-1)]) | [0,0] | [0,1] | [1,0] | [1,1] |
| [0,0] | 0 | 0 | 1 | 1 |
| [0,1] | 1 | 1 | 0 | 0 |
| [1,0] | 0 | 1 | 1 | 0 |
| [1,1] | 1 | 0 | 0 | 1 |

Table 6. Truth table generating the output bit *O(j)*.

From Table 6 one observes that the output *O(j)* will have a *0.5* probability of being a *0* or *1* bit whenever the probabilities of the vectors *B$_Y$(j-1)* , *B$_Y$(j-2)* y *B(j) = [B$_X$(j);B$_Y$(j)]* are also *0.5*. And we can say that the distribution of  *0*'s and *1*'s the output binary sequence will be uniform as long as the distribution of the sequence of the binary vector *B(j) = [B$_X$(j);B$_Y$(j)]* is uniform.

But *B(j)* is obtained in the output of the sub-space division block, as *t=j=P*i*, *B$_X$(j)= b$^{P*i}_X$* and *B$_Y$(j)= b$^{P*i}_Y$,* and as it was seen the pattern of visits of any sub-space is not uniform, it depends of the dynamic of the chaotic attractor. The dynamic of visits of every sub-space is analyzed in Fig. 7.

In Fig.7, it is observed that the sequences of binary vectors *[b$^i_X$ , b$^i_Y$]* in (a) and *[b$^{P*i}_X$ , b$^{P*i}_Y$]* in (b) have a characteristic evolution. In particular, from the Fig.7 (a) is easy to conclude that it is possible to predict next states in the automata given and initial state for *P=1*. For example *P([b$^i_X$ , b$^i_Y$] =[0,0])/([b$^{i-1}_X$ , b$^{i-1}_Y$] = [1,0])=1*. On the other hand, in Fig.7 (b) this sequence is sampled at *P* rate, when *P* is taken sufficiently high, *P>>1*. Then, the correlation of the states in the chaotic system disappears and any transition is possible.

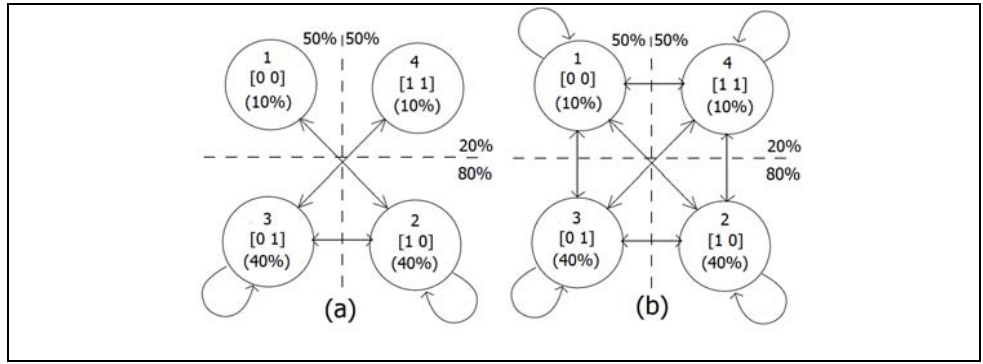

Fig. 7. Finite automata for the evolution dynamics of binary vector $[b^{P*i}{}_X, b^{P*i}{}_Y]$ generated in the chaotic PRBG with system A (a) when $P=1$ and (b) when $P>>1$.

To measure the predictability, different binary sequences are produced in the chaotic PRBG with initial conditions $S2$ of Table 5, a length of $n=2Mill$ of bits and different values of sampling factor $P=1, 101$ y $151$. With these sequences, it is measured the probability of transition from a state $B(j-1) = [B_X(j-1), B_Y(j-1)]$ at instant $t=j-1$ to the state $B(j) = [B_X(j), B_Y(j)]$ in the next instant. These measurements can be seen in Table 7.

**NUMBER OF VISITS OF EVERY SUBSPACE (from B[j-1] to B[j]) (P=1)**

| 2000000 | 211026 | 788125 | 789279 | 211570 | TOTAL |
|---|---|---|---|---|---|
| B(j-1) \ B(j) | [0,0] | [1,0] | [0,1] | [1,1] | 2000000 |
| [0,0] | 0 | 211025 | 0 | 0 | 211025 |
| [1,0] | 211026 | 104267 | 472832 | 0 | 788125 |
| [0,1] | 0 | 472833 | 104877 | 211570 | 789280 |
| [1,1] | 0 | 0 | 211570 | 0 | 211570 |

(a)

**PROBABILITY OF VISITS OF EVERY SUBSPACE (from B[j-1] to B[j]) (P=1)**

| TOTAL | 0,105513 | 0,394062 | 0,394639 | 0,105785 |
|---|---|---|---|---|
| t=j-1 \ t=j | [0,0] | [1,0] | [0,1] | [1,1] |
| [0,0] | 0 | 1 | 0 | 0 |
| [1,0] | 0.267757 | 0.132298 | 0.599945 | 0 |
| [0,1] | 0 | 0.599069 | 0.132877 | 0.268054 |
| [1,1] | 0 | 0 | 1 | 0 |

(b)

**NUMBER OF VISITS OF EVERY SUBSPACE (from B[j-1] to B[j]) (P=101)**

| 2000000 | 210811 | 788153 | 789383 | 211653 | TOTAL |
|---|---|---|---|---|---|
| B(j-1) \ B(j) | [0,0] | [1,0] | [0,1] | [1,1] | 2000000 |
| [0,0] | 14243 | 91059 | 90979 | 14530 | 210811 |
| [1,0] | 91236 | 302043 | 303811 | 91063 | 788153 |
| [0,1] | 90827 | 303940 | 302989 | 91627 | 789383 |
| [1,1] | 14505 | 91111 | 91604 | 14433 | 211653 |

(c)

**PROBABILITY OF VISITS OF EVERY SUBSPACE (from B[j-1] to B[j]) (P=101)**

| TOTAL | 0,105406 | 0,394077 | 0,394691 | 0,105827 |
|---|---|---|---|---|
| t=j-1 \ t=j | [0,0] | [1,0] | [0,1] | [1,1] |
| [0,0] | 0.0675629 | 0.431946 | 0.431567 | 0.0689243 |
| [1,0] | 0.115759 | 0.383229 | 0.385472 | 0.11554 |
| [0,1] | 0.115061 | 0.385035 | 0.38383 | 0.116074 |
| [1,1] | 0.068532 | 0.430473 | 0.432803 | 0.0681918 |

(d)

**NUMBER OF VISITS OF EVERY SUBSPACE (from B[j-1] to B[j]) (P=151)**

| 2000000 | 211841 | 788804 | 788004 | 211351 | TOTAL |
|---|---|---|---|---|---|
| B(j-1) \ B(j) | [0,0] | [1,0] | [0,1] | [1,1] | 2000000 |
| [0,0] | 24757 | 81217 | 81281 | 24586 | 211841 |
| [1,0] | 81133 | 313633 | 313237 | 80800 | 788803 |
| [0,1] | 81244 | 312899 | 312659 | 81203 | 788005 |
| [1,1] | 24707 | 81055 | 80827 | 24762 | 211351 |

(e)

**PROBABILITY OF VISITS OF EVERY SUBSPACE (from B[j-1] to B[j]) (P=151)**

| TOTAL | 0,105921 | 0,394402 | 0,394002 | 0,105676 |
|---|---|---|---|---|
| t=j-1 \ t=j | [0,0] | [1,0] | [0,1] | [1,1] |
| [0,0] | 0.116866 | 0.383387 | 0.383689 | 0.116059 |
| [1,0] | 0.102856 | 0.397606 | 0.397104 | 0.102434 |
| [0,1] | 0.103101 | 0.397077 | 0.396773 | 0.103049 |
| [1,1] | 0.1169 | 0.383509 | 0.38243 | 0.117161 |

(f)

Table 7. Measurements of the probability of transition from one subspace to another.

Table 7 shows two tables for every value of *P* used in the measurements. The table on the left counts the visits or transitions from the sub-space where $B(j-1) = [B_X(j-1),B_Y(j-1)]$ was at instant $t=j-1$ to the sub-space where $[B_X(j),B_Y(j)]$ is at the following instant. The table on the right calculates the probability of transition between the four subspaces calculated from the table on the left. The values in the different tables are obtained in (a), (b) with *P=1*, in (c), (d) *P=101*, and in (e), (f) *P= 151*. As a conclusion of these, it is found that the system at instant *t=j* has the following probabilities for being in any of the four possible states:

$$P([B_X(j),B_Y(j)]=[0,0])= P(([B_X(j),B_Y(j)]=[1,1])= 0.1 \qquad (10)$$

$$P([B_X(j),B_Y(j)]=[0,1])= P(([B_X(j),B_Y(j)]=[1,0])= 0.4$$

It is observed that the greater the value of *P,* the closer the probabilities to *0.5*. The case where *P=151* is then considered to proceed with the measurements. Now it is possible to calculate from Table 7 the probability of obtaining $B_X(j) = 0$ or $B_X(j)=1$ from a given predecessor state $B(j-1) = [B_X(j-1),B_Y(j-1)]$. And the same for the probabilities of $B_Y(j) = 0$ or $B_Y(j)=1$ from the state $B(j-1)$. These calculus can be done by the reader in all cases with the data of Table 7(f), but here the first one is illustrated as an example:

$$P(B_X(j) = 0/B(j-1)=[0,0]) = 0{,}116866* P(B(j-1)=[0,0])+ 0{,}383689* P(B(j-1)=[0,0])=$$
$$0{,}500555* P(B(j-1)=[0,0]) \qquad (11)$$

All results are summarized in Table 8, where $\Delta\varepsilon$ is the deviasion from probability *0.5*:

| P=151 t=j-1 \ t=j | By=0 | By=1 | Δε | Δε |
|---|---|---|---|---|
| [0,0] | 0.500253 | 0.499748 | 0.000253 | 0.000252 |
| [1,0] | 0.500462 | 0.499538 | 0.000462 | 0.000462 |
| [0,1] | 0.500178 | 0.499822 | 0.000178 | 0.000178 |
| [1,1] | 0.500409 | 0.499591 | 0.000409 | 0.000409 |

(a)

| P=151 t=j-1 \ t=j | Bx=0 | Bx=1 | Δε | Δε |
|---|---|---|---|---|
| [0,0] | 0,500555 | 0,499446 | 0,000555 | 0,000554 |
| [1,0] | 0,49996 | 0,50004 | 0,000040 | 0,000040 |
| [0,1] | 0,499874 | 0,500126 | 0,000126 | 0,000126 |
| [1,1] | 0,49933 | 0,50067 | 0,00067 | 0,00067 |

(b)

Table 8. Calculus of the probability of obtaining from the estate $B(j-1) = [B_X(j-1),B_Y(j-1)]$ (a) $B_X(j) = 0$ or $B_X(j)=1$ and (b) $B_Y(j) = 0$ or $B_Y(j)=1$.

From Table 8 it can be seen that:

$$P(B_X(j) = 0) \approx P(B_X(j) = 1) \approx P(B_Y(j) = 0) \approx P(B_Y(j) = 1) = 0.5 \pm \Delta\varepsilon \qquad (12)$$

The error $\Delta\varepsilon$ is different in each case of Eq. (12). To get this equation we have taken the approximations of Eq. (13) for all cases:

$$P(B_X(j) = 0) \approx 0.5 * (P([B_X(j-1)/B_Y(j-1)=[0,0])+ P([B_X(j-1)/B_Y(j-1)=[0,1])+$$
$$P([B_X(j-1)/B_Y(j-1)=[1,0])+ P([B_X(j-1)/B_Y(j-1)=[1,1])) \approx 0.5 \pm \Delta\varepsilon \qquad (13)$$

Considering all cases, we obtain that $\Delta\varepsilon<0.0006$. This result can be compared with the the acuracy of this measurement $\delta=1/n= 5x10^{-7}$, where *n* is the length of the sequence. Finally it can be concluded that the distribution of the binary vector *B(j)* is uniform with an error of $\Delta\varepsilon_{max}<0.0006$. And from Table 8 we conclude that the probability of producing a bit *0* or *1* given a sequence *S2* of *n=2Mill* bits is almost *0.5* within a maximum error of $\Delta\varepsilon_{max}<0.0006$. This gives a quantitative magnitude of the chaotic PRBG's predictability.

d) Performance Test

To assess the functionality of the PRBG for its use in cryptography, it is necessary to measure its speed. As it is said, an allegedly "secure" cipher is not acceptable for real applications if it is not efficient.

To establish the complexity, and consequently the performance of the chaotic PRBG, the principle of invariance is observed. This says that the efficiency of one algorithm in different execution environments differs only in a multiplicative constant, when the values of the parameters of cost are sufficiently high. In this case, the asymptotic behavior of the computational cost of the PRBG is governed by the calculus performed in the chaotic block of Fig.3. This block makes $P$ iterations for each output bit, $O(j)$. The capital theta notation ($\Theta$) can be used to describe an asymptotic tight bound for the magnitude of cost of the PRBG. And consequently, the 2D symmetric coupled logistic maps have a computational cost or complexity of order $\Theta(P*n)$. This result is typical in chaotic based cryptosystems where multiple iterations are required for each encryption step. In this case the number of iterations required for good statistical quality is around *100* iterations and this can be considered small compared with other cryptosystems ( Examples: larger than *250* or *65536* in (Kocarev, 2001)). Nevertheless in our case, the calculus is two dimensional and two real variables are computed for each step.

To measure the speed of the chaotic PRBG the following test is performed: the algorithm is implemented in C++ code with the compiler **Bloodshed C++ 4.9.9.2** and run in PC with **Windows 2000 5.00.2195 SP4 - Intel Pentium 4 - CPU 2.80GHz – 515.056 KB RAM**. Generally speaking, on a PC with a $f_z$ Hz CPU, it is acceptable if the encryption speed is $f_z/a$ bps, where *a <=100* (Alvarez & Li, 2006). In our case the chaotic PRBG gives a speed of *a=2394* if *P=25* and *a=4219* when *P=47.*

The results obtained show no optimal performance, but as it is known (Alvarez & Li, 2006) besides the CPU frequency, the encryption speed of a software implementation is tightly dependent on other issues, such as the CPU structure, the memory size, the underlying OS platform, the developing language, all options of the compiler, and so on. In consequence one could say that this speed needs to be increased with code optimization. In this case, among other things, it could be convenient to use hardware implementations with parallel mechanisms for calculating simultaneously each variable of the 2D chaotic iterations.

### 3.3 Second example: Improving encryption with symmetry swap of chaotic variables

In this example it is explored how the geometric and symmetric characteristics of the chaotic maps can be used to improve the encryption algorithm. In particular, these characteristics are studied in the chaotic logistic bimap attractors and a mechanism called "***symmetry-swap***" (Pellicer-Lostao & López-Ruiz, 2009) is introduced to enhance the previous PRBG. It is shown that this mechanism can increase the degrees of freedom of the key space, while maintaining the performance of the PRBG algorithm.

Let us observe that the maps under consideration present symmetry with respect to the diagonal axis. Now let us consider a point *[$x_0$, $y_0$]* and its conjugated *[$y_0$, $x_0$]* with respect to the diagonal *[$x_0$, $y_0$]* as two different initial conditions. It can be seen, that starting from any of these points systems A and B produce different but symmetric orbits (conjugated orbits).

Now it becomes clear that an interchange (or swap) of coordinates $x$ and $y$ in an orbit state will to produce a jump to a conjugated orbit, while the attractor and the chaotic regime are not affected. These facts are illustrated in the schematic diagram of Fig. 8 for Systems A and B. When starting with the same initial conditions, one orbit and its conjugated are presented, jumping from one to the other is possible thanks to the swapping factor.

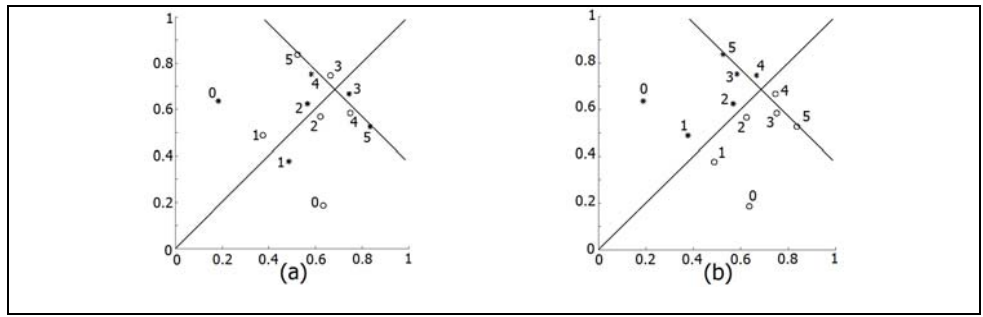

Fig. 8. Five points in the evolution of trajectories for System A (a) or B (b) when starting with $\lambda = 1.07$ and the Same Initial Conditions $[x_0, y_0]$ (point **0** with the star-marker) and its Coordinate Conjugated $[y_0, x_0]$ (point **0** with the circle-marker).

In these circumstances, a swapping of coordinates could be introduced in the algorithm of Fig. 3, without altering its pseudorandom properties. This mechanism is named by the authors as "***symmetry-swap***". In practice, the s symmetry-swap can be an additional step at the input of the system, which is applied at specific instants $t=i-1$ as desired. When the swapping is applied at a constant rate $S$, a swap of coordinates is introduced every $S$ iterations and the algorithm has the following performance: the map evolves along one specific orbit during $S$ iterations and after a swap in the coordinates (swapping $x \leftrightarrow y$), the map jumps to a conjugated orbit. Let us call $S$ the rate of swapping or the swapping factor.

a) Calculus of the key space

It is important to observe at this point, that the introduction of a swapping factor S does not penalizes the computational cost of the resulting PRBG. The chaotic block again dominates its asymptotic behavior. As a result, the swapped 2D symmetric coupled logistic maps PRBGs have an asymptotic tight bound of order $\Theta(P*n)$. Then we must expect a similar performance as for the algorithm in the first example. Another valuable aspect to remark is that the swapping factor $S$ can offer an improvement in the range of input values of the initial PRBG algorithm. In cryptography, this means an enhancement in security and it can be obtained straight from the fact that $S$, considered as a constant value, may represent a new free parameter in the key-space.

Let us consider that the useful values of $S$ could range in the interval $S = [1,n]$, where $n$ is the number of bits generated. Taking $n$ for a typical value of *1 Mill.* of bits, this would enlarge the key space calculated last example. Following analogous calculations and $S$ taking integer values, then *[S] =1000000*, $K = K = K_{\lambda X} \times K_{\lambda Y} \times K_{x0} \times K_{y0} \times K_P \times K_S$. This increases the key size from *93* to *113* for single precision and from *209* to *229* for double. The enlargement of the key space makes the swapped algorithm stronger against brute force attack than the non swapped one.

b) Statistical testing

The interesting point about the symmetry-swap, is that no matter what number of consecutive iterations and swaps are performed to the system, the chaotic behavior always prevails. Logically, this particular fact will make pseudorandomness to prevail too.

The authors explore the construction of a swapped PRBG. this is, following algorithm in Fig. 3 with system A and adding a constant swapping factor of value *S* in the input. Ten pseudo-random binary sequences are generated with the same characteristics (initial conditions and length) as the ones described in Table 5. A swapping factor of *S = 90* is applied for sequences to be tested with Diehard test suite. To illustrate a different value, a swapping factor of *S = 50* is chosen with NIST's suite. The sequences of the swapped PRBGs demonstrate similar random results when submitted to the tests. Very similar $P_{min}$ values to those in Table 5, or even the same, are obtained in all cases. Fig. 9 shows graphically the results obtained and illustrated the success of the tests. Unsurprisingly, this demonstrates that in this case the symmetry-swap maintains pseudo randomness.

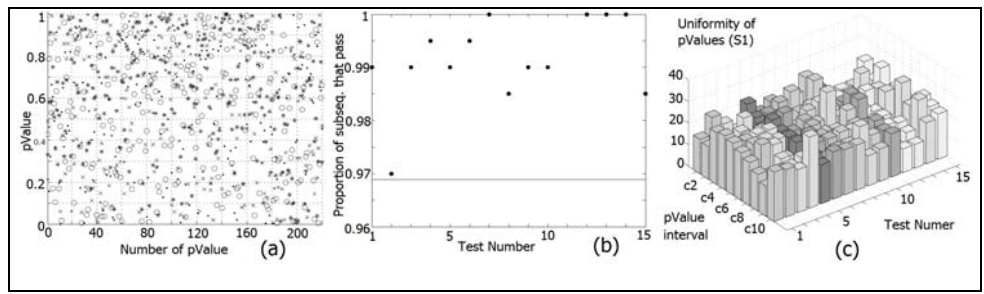

Fig. 9. In (a), *p-values* obtained with all tests of Diehard suite for initial conditions *S1, S2, S3* and S4 with $P = P_{Dmin}$ and *S = 90*. In (b) the proportion of sub-sequences that pass NIST test suite with initial conditions *S1, P =$P_{Nmin}$* and *S = 50*. In (c) The distribution of p-values for each test with the same conditions of (b) demonstrates the required uniformity.

In addition, one may think that the introduction of a swapping factor *S* can be applied in multiple ways. Consider, for example, different values of *S* used alternatively in the process, this may make the swapping factor many dimensional. Another way could be to consider an *S* value variable in time. The swapping factor can also offer an easy feedback mechanism, when making its value dependable of the output. Therefore the symmetry-swap mechanism is a very flexible tool. In the end, it can be observed that the symmetry-swap offers a remarkable advantage, while maintaining speed and simplicity of the initial PRBG algorithm.

To conclude, the symmetry swap mechanism represents an interesting strategy for finding additional degrees of freedom in the key space with no extra computational costs. This is possible thanks to the symmetric and geometric properties observed in the chaotic attractor. Unlike what happens when introducing the sampling factor *P*, the swapping factor *S* doesn't force the designer to consider a trade-off between its range of values and the speed of the algorithm. Moreover this degree of freedom can be introduced in multiple ways. Therefore the swapping factor *S* can increase the security of the system with great flexibility.

## 4. Conclusion and perspectives

Chaotic cryptography has experienced significant development since its birth in the early 90's. Nevertheless, as other interdisciplinary areas of research, it has encountered many difficulties in its way through. Chaotic cryptographers have been censured by conventional cryptographers of a lack of deep knowledge in cryptography.

Chaotic cryptography appears as an applied branch of non-linear dynamical systems. Initial works were discussed in the area of applied cryptology (Dachselt & Schwarz, 2001; Amigó, 2009) but soon demonstrated low security or performance, being discarded before long. Since then, numerous cryptosystems have been proposed, many still lacking of thorough security and efficiency analysis or difficult to implement in practice. This has damaged the image of chaotic cryptography in the cryptographic community.

Nevertheless hard work has been done to this respect (Alvarez & Li, 2006) and relevant aspects have been identified as important in the process of designing good chaotic cryptosystems. As a consequence, there are a series of guidelines to assist designers to fulfil some basic cryptographic requirements (Alvarez & Li, 2006). They address three main issues: implementation, key management and security analysis. Implementation guidelines take into account the necessity of a detailed description of the cryptosystem and measurements of its performance. Regarding to the key, a complete analysis should be done to guaranty the required behaviour of the chaotic algorithm in the key space. This means looking for regions of weak keys or windows of non chaotic behaviour, among others. The definition of the key and the procedure to build it, should be also clearly specified.

As for the security analysis, the novelty of chaotic cryptography makes difficult a rigorous study on the security of these algorithms (Amigó, 2009). As well as it is not yet established a general analytical methodology to study non-linear dynamical systems, there is also a lack of adequate analysis tools to assess the behaviour of these systems as part of the cryptographic algorithm. In this case, conventional cryptography tools cannot be directly applied to chaotic cryptographic algorithms. This is another reason why these totally new schemes seem doubtful to conventional cryptographers.

However, lessons have been learnt and the work done in this area (Alvarez et al, 2011) is giving a series of mathematical tools to assess the security of the algorithms. Chaotic cryptoanalysis works show examples of the tools used to assess security and the details of their realization. The analysis of the dynamical properties of the systems as the bifurcation diagrams and the Lyapunov's exponents is considered a basic security analysis. These tools allow the designer to discover regions of the parameters of design where the chaotic system shows windows of periodic behaviour. Furthermore a study of the statistical characteristics of the cryptosystem must be recommended. This is done, through the study of the histograms looking for uniformity in the distribution of ciphertexts and calculating for different values of the parameters the entropy of the orbits and the statistical complexity.

Today chaotic cryptography remains a very active field, with many publications in different areas of modern cryptography. This interdisciplinary approach has been quite successful in pseudo random number generation. But it is still in the side of chaotic cryptography to develop the necessary analytical tools and prove the compliance of the required standards.

The chapter concludes here. It has intended to give a vision of what chaotic cryptography is and a comprehensive example of the techniques used in this field. This example gathers a series of topics considered of interest in the field of chaotic cryptography, such as digital chaotic cryptosystems, chaos based Pseudo-Random Bit Generation (PRBG) and the use of chaotic two-dimensional maps. It also covers a detailed explanation of its implementation and describes the measurements performed to assess its performance characteristics. This provides a display of how digital chaotic cryptosystems are built and the tools used in this area to evaluate their quality. Additionally it is presented a mechanism that illustrates how the geometry characteristics of the chaotic attractors can be used to improve the security of the cryptographic algorithm.

The authors would like to leave a final remark. This is that, though chaotic cryptography may be considered at present peripheral in circles of conventional crypto, chaotic number generation may have attractive applications as simulation engines in computational science (Pellicer-Lostao & Lopez-Ruiz, (2011c)). Chaos based number generators are easy to use and highly configurable. This makes them a valuable tool for this application.

## 4. Acknowledgment

The authors thank the BIFI Institute of the Universidad de Zaragoza by financial support.